\documentclass[showpacs,preprint,preprintnumbers,showpacs,showkeys,superscriptaddress,amsmath,amssymb,nofootinbib]{revtex4-2}
\usepackage{bbold}
\usepackage{color}
\usepackage{latexsym}
\usepackage{amsmath}
\usepackage{amssymb}
\usepackage[utf8]{inputenc}
\usepackage{amsmath}
\usepackage{amsfonts}
\usepackage{bm}
\usepackage{bbold}
\usepackage{amssymb}
\usepackage{epstopdf}
\usepackage{epsfig}
\usepackage{eufrak}
\usepackage{euscript}
\usepackage{pstricks}
\usepackage{graphics}
\usepackage{graphicx}
\usepackage{picture}
\usepackage{appendix}
\usepackage{enumitem}

\newcommand{\be}{\begin{equation}}
\newcommand{\ee}{\end{equation}}
\newcommand{\ba}{\begin{eqnarray}}
\newcommand{\ea}{\end{eqnarray}}

\begin{document}
\title{\Large{Probing the interference between non-linear, axionic and space-time-anisotropy effects
in the QED vacuum}}
\author{ J. M. A. Paix\~ao} \email{jeff@cbpf.br}
\affiliation{Centro Brasileiro de Pesquisas F\'isicas, Rua Dr. Xavier Sigaud
150, Urca, CEP 22290-180, Rio de Janeiro, Brazil}
\author{ L. P. R. Ospedal } \email{leoopr@cbpf.br}
\affiliation{Centro Brasileiro de Pesquisas F\'isicas, Rua Dr. Xavier Sigaud
150, Urca, CEP 22290-180, Rio de Janeiro, Brazil}
\author{M. J. Neves}\email{mariojr@ufrrj.br}
%\affiliation{Department of Physics and Astronomy, University of Alabama, Tuscaloosa, Alabama 35487, USA}
\affiliation{Departamento de F\'isica, Universidade Federal Rural do Rio de Janeiro, BR 465-07, CEP 23890-971, Serop\'edica, RJ, Brazil}
\author{J. A.  Helay\"el-Neto}\email{helayel@cbpf.br}
\affiliation{Centro Brasileiro de Pesquisas F\'isicas, Rua Dr. Xavier Sigaud
150, Urca, CEP 22290-180, Rio de Janeiro, Brazil}
%

%\date{\today}

%\pacs{04.50.-h, 05.20.-y, 05.90.+m}

%\vspace{-5cm}

%\arxivnumber{....arxiv:}

\begin{abstract}
%\noindent
%

%
In this paper, we pursue the investigation of a generic non-linear extension of axionic electrodynamics in a Carroll-Field-Jackiw (CFJ) scenario that implements Lorentz-symmetry violation (LSV). The model we inspect consists of an arbitrary non-linear electrodynamic action coupled to the axion field in presence of an anisotropy four-vector that realizes the breaking of Lorentz symmetry under the particle point of view. For the sake of our considerations, the non-linear electromagnetic field is expanded around a constant and uniform magnetic background up to second order in the propagating photon field. The focus of our attention is the study of the material properties of the vacuum in the particular case of a space-like CFJ $4$-vector. The dispersion relations associated to the plane wave solutions are explicitly worked out in two situations: the magnetic background perpendicular and parallel to the wave direction. We extend these results to consider the analysis of the birefringence phenomenon in presence of non-linearity, the axion and the LSV
manifested through the spatial anisotropy. Three specific proposals of non-linear electrodynamics are contemplated: Euler-Heisenberg (EH), Born-Infeld (BI) and the Modified Maxwell electrodynamics (ModMax). Throughout the paper, we shall justify why we follow the unusual path of connecting, in a single Lagrangian density, three pieces of physics beyond the Standard Model, namely, non-linearity, axions and LSV. We anticipate that we shall not be claiming that the simultaneous introduction of these three topics beyond the Standard Model will bring new insights or clues for the efforts to detect axions or to constrain parameters associate to both non-linear electrodynamics and LSV physics . Our true goal is to actually inspect and describe how axionic, non-linear and LSV effects interfere with one another whenever physical entities like group velocity, refraction indices, birefringence and effective masses of physical excitations are computed in presence of an external constant and homogeneous magnetic field.

\end{abstract}
%

%\pacs{11.15.-q,12.60.-i,11.10.Lm}
%11.15.−q Gauge field theories
%12.60.−i Models beyond the standard model
%11.10.Lm Nonlinear or nonlocal theories and models

%
%\keywords{ Non-linear electrodynamics, Axion-like particle, Dispersion relations.}

\maketitle

\newpage

\pagestyle{myheadings}
%
%\markright{Probing the interference between non-linear, axionic and space-time-anisotropy effects
%in the QED vacuum}
%
%
\section{Introduction}
\label{sec1}
The strong CP problem is still an intriguing question in the Standard Model (SM) of elementary particles. Certainly, the mechanism proposed by Peccei and Quinn is the most popular and elegant approach to solve this issue by introducing the axions \cite{Peccei,Peccei2}. We should point out here that there are, however, other resolutions to solve the strong CP problem without introducing new particles beyond the Standard Model, as one may find in the articles of Refs. \cite{Wen,Nakamura,Nodoka,Nodoka2}. The Axion-like Particles (ALPs) has been the subject of investigation in diverse branches of the high energy physics. A good motivation is that such particles are strong candidates for the dark matter content \cite{Wilczekaxion,Sikivie,Dine}. Furthermore, ALPs naturally arise in string theories \cite{axionstring}.
Over the past decades, a considerable effort has been made for the detection of ALPs, both in astrophysical experiments \cite{Ayala,Reynolds,Bondarenko,CAST,jacobsen2022constraining} and in particle accelerators \cite{Baldenegro1,Baldenegro2,d'Enterria,SCHOEFFEL}. The challenge is that the ALPs couple very weakly to the SM matter, so the bounds obtained have a stringent parameter space. For example, the CAST experiment that searches for ALPs produced in the solar core provides a well-established limit for the ALP-photon interaction with coupling constant $g_{a\gamma} \simeq 0.66 \times 10^{-10} \, \mbox{GeV}^{-1}$ and ALP mass restricted to $m_a < 0.02 \, \mbox{eV}$ at 95\% CL \cite{CAST}. We also highlight that ALPs can be produced by ALP-photon conversion in the presence of an intense magnetic  background field, as described by the Primakoff process. There is also a great deal of interest in looking for ALPs with very small masses, the so-called ultra-light Axions (ULAs). They are particles with mass in the broad range of $10^{-33} \, \mathrm{eV} \lesssim m_a \lesssim 10^{-18} \, \mathrm{eV}$. The search for ULAs involves cosmological observations from the cosmic microwave background (CMB) and the large-scale structure (LSS). A thorough review is provided in ref. \cite{marsh}.

In presence of intense magnetic fields close to the Schwinger's critical magnetic field, {\it i.e.}, $|{\bf B}|_S = m_e^2/q_e = 4.41 \times 10^{9}\,\mbox{T}$, non-linear effects acquire relevance \cite{Schwinger}. In the work \cite{nossoJHEP}, a general approach has been followed to investigate ALPs in non-linear electrodynamic scenarios, where some optical properties of the vacuum have been investigated, such as the vacuum magnetic birefringence (VMB) and  Kerr effect. Furthermore, it has been shown that the presence of the axion generates dispersion relations that depend on the wavelength, so that dispersive refractive indices show up that would not be present if only non-linearity were considered. In a seminal work \cite{Raffelt_1988}, the authors connected axionic physics with the Euler-Heisenberg electrodynamics and discussed birefringence experiments, photon-axion conversion, as well as the axion-graviton conversion in the vicinity of stars with an intense magnetic fields.
It is well-known that the ALP-photon conversion in a magnetic background changes the optical properties of the vacuum. Therefore, the measure of the VMB can provide bounds on the axion mass and coupling constant $g_{a\gamma}$ \cite{Maiani,Yamazaki,Villalba}. Although VMB is an effect predicted by quantum electrodynamics (QED), there is still no experimental evidence of its existence  produced in laboratories. The PVLAS Collaboration was one of the most notable projects in this search, having ended its activity in 2017 after 25 years of efforts to measure the birefringence and vacuum dichroism phenomena, providing very reliable limits for such quantities  \cite{25years,Zavattini,DellaValle}. Even so, indirect evidence of vacuum birefringence was found from measurement of optical polarization of the neutron star RX J1856.5-3754 \cite{Mignani}.
At this stage, it is worthy mentioning that the axionic interaction term can be generated via radiative corrections in a theory with Lorentz symmetry violation (LSV) \cite{borges}. Although Lorentz invariance is a fundamental invariance principle in elementary particle physics and in Einstein's classical General Relativity, it is known that, for a quantum theory of gravity, such invariance may not hold any longer. For example, in string theories \cite{LSVstring theory,coleman,Mavromatos1,Mavromatos2,moffat}, it is estimated that there should be small violations of Lorentz symmetry next to the Planck energy scale, namely, $E_{Pl} = 10^{19} \, \mathrm{GeV}$. This would occur in the early universe. In this perspective, we are motivated to study which influence a LSV background would have on an axionic theory. To generalize our approach, we consider that the electrodynamics model may have non-linear contributions. Although, in specific cases, one can easily reduce to the usual Maxwell's  theory. Moreover, we point out that LSV theories introduce an anisotropy in space-time, such that  is reasonable to obtain  a birefringence effect \cite{Lorentz-violation-birefringence,Kostelecky}. This characteristic added to the non-linear ALP-photon mixing model \cite{nossoJHEP} generates a very rich effective model with implications in the optical properties of the vacuum. In particular, for the LSV term, we adopt the Carroll-Field-Jackiw (CFJ) electrodynamics \cite{cfj}, which is a generalization of a Chern-Simons term for $(3+1)$ dimensions. For the consistency of the model, a quadrivector is introduced that guarantees the gauge symmetry of the theory, but does not preserve the Lorentz and CPT symmetry. The CFJ term appears in the CPT-odd gauge sector of the Standard Model Extension (SME). This model developed by Colladay and Kosteletsky describes a general action with terms that violate Lorentz and CPT symmetry \cite{Colladay1,Colladay2}.
There is a rich literature on the CFJ electrodynamic model. In the work \cite{yuri}, limits were obtained for the CFJ Lorentz-breaking parameter in the time-like case through laboratory experiments such as quantum corrections to the spectrum of the hydrogen atom, electric dipole moment, as well as the interparticle potential between fermions. Studies on the possible effects of contributions of the CFJ model for the cosmic microwave background (CMB) were carried out in ref. \cite{CMB}. Recently, in the supersymmetry scenario, the gauge boson-gaugino mixing was investigated by taking into account the effects of the LSV due to a CFJ term \cite{cfjsupersimetria}. We also highlight that there is a connection between the axionic theory and Lorentz and CPT-violation. For example, in ref.\cite{Yakov}, the author establishes this connection by embedding Carroll-Field-Jackiw (CFJ) electrodynamics in a premetric framework. The fact is used that in CFJ electrodynamics the constraint $\partial_\mu v_\nu -\partial_\nu v_\mu =0$ allows writing the Lorentz-breaking vector as the gradient of a scalar $v_\mu = \partial_\mu \phi$; so, in performing this redefinition in the pre-metric Lagrangian, the axionic interaction term naturally arises. However, this approach does not provide a dynamics for the axion. Also, the corresponding energy-momentum tensor does not depend on the axion field. Furthermore, the author discussed the relation between the birefringence phenomenon with Lorentz and CPT symmetry violation.  It is possible to associate the non-observation of birefringence with the preservation of these symmetries. For more details on LSV, we indicate the review \cite{Kostelecky_Russell} and references therein.

Before going on and starting to work out the developments of our paper, we would like to call into question our motivation to bring together three different physical scenarios beyond the Standard Model in a single Lagrangian, namely: axions, non-linear electrodynamic extensions and Lorentz-symmetry violating physics (LSV is here realized by means of the Carroll-Field-Jackiw term). The usual procedure is to consider each of these physical situations separately, once we expect that their respective individual effects correspond to tiny corrections to current physics. Connecting these three diverse physics in a single action might appear as a waste of efforts or, simply, an exercise to mix up different effects. Nevertheless, what we truly wish by coupling axions to non-linear electrodynamics and LSV physics is to show how the parameters associated to the axion and LSV sectors couple to external electric and magnetic fields whenever non-linearity is considered. Actually, the main effort we endeavor is to inspect how the magnetic background field may broaden the effects of the tiny axionic and LSV parameters on physical properties such as birefringence, refractive indices, dichroism and group velocity. This is investigated with the help of the dispersion relations we shall derive in different situations characterized by particular configurations of external fields.
And to enforce our claim to consider the simultaneous presence of these three sorts of effects, we gather some works in Refs. \cite{Keser,Sekine,Li,Lehnert}. In these papers, non-linear quantum electrodynamics, axion electrodynamics and LSV are studied in Condensed Matter scenarios such as Dirac and Weyl semimetals and topological magnetic materials. We are then motivated to assume that topological materials appear as a natural laboratory that justify the inspection of how the effects of non-linearity, axions and LSV interfere with each other. Cosmology provides another viable scenario that may justify efforts in the quest for the interference between the three effects we are here discussing. In Refs. \cite{novello,phe,marsh}, we cast
reference works that support our proposal.
Finally, knowing that non-linearity, axions and LSV are issues currently investigated in connection with astrophysical structures \cite{denisov,galanti,hli}, we can also elect Astrophysics as another field of interest to study the concomitant presence of these three issues and how they affect each other in the study of the propagation of electromagnetic waves in the QED vacuum.

%
%Before going on and starting to work out the developments of our paper, we would like to call into question our motivation to bring together three different physical scenarios beyond the Standard Model in a single Lagrangian, namely: axions, non-linear electrodynamic extensions and Lorentz-symmetry violating physics (LSV is here realized by means of the Carroll-Field-Jackiw term). The usual procedure is to consider each of these physical situations separately, once we expect that their respective individual effects correspond to tiny corrections to current physics. Connecting these three diverse physics in a single action might appear as a waste of efforts or, simply, an exercise to mix up different effects. Nevertheless, what we truly wish by coupling axions to non-linear electrodynamics and LSV physics is to show how the parameters associated to the axion and LSV sectors couple to external electric and magnetic fields approached by introducing non-linearity. Actually, the main effort we endeavor is to inspect to which extent strong external electric and magnetic fields may broaden the effects of the tiny axionic and LSV parameters on physical properties such as birefringence, refractive indices, dichroism and group velocity. This is investigated with the help of the dispersion relations we shall derive in different situations characterized by particular configurations of external fields.

%
In this contribution, we investigate the propagation effects of a general axionic non-linear ED in presence of a CFJ term. As mentioned, the CFJ introduces the
$4$-vector that breaks the Lorentz symmetry, and the isotropy of the space-time. We introduce a uniform magnetic field expanding the propagating field of the model up to second order around this background field. The properties of the medium are discussed in presence of the magnetic background. We obtain the dispersion relations of the linearized theory in terms of the magnetic background, the CFJ $4$-vector, and the axion coupling constant.  The case of a space-like  quadrivector is analysed, such that the plane wave frequencies are functions of the wave vector $({\bf k})$,  the magnetic background $({\bf B})$, and the CFJ background vector $({\bf v})$. Thereby, we consider two cases : (a) when ${\bf k}$, ${\bf B}$ and ${\bf v}$ are perpendiculars, and (b) when ${\bf k}$ is parallel to ${\bf B}$, but both vectors remain perpendicular to ${\bf v}$. The solutions of these cases define the perpendicular and parallel frequencies, respectively. Using these dispersion relations, we calculate the birefringence through the perpendicular and parallel refractive indices. We apply  our results to the non-linear electrodynamics of Euler-Heisenberg \cite{EulerHeisen}, Born-Infeld \cite{BornInfeld}, and Modified Maxwell (ModMax) \cite{Bandos,Sorokin,Sorokin2}.

This paper is organized according to the following outline: In Section (\ref{sec2}), the axionic non-linear theory is presented with the CFJ term in an electromagnetic background field. In Section (\ref{sec3}), we consider a purely magnetic background field and obtain the permittivity and permeability tensors, as well as the dispersion relations associated with the plane wave solutions. Next, in Section (\ref{sec4}), the birefringence phenomenon is discussed in the framework of Euler-Heisenberg,  Born-Infeld, and  ModMax electrodynamics. Finally, the Conclusions and Perspectives are cast in Section (\ref{sec5}).
We adopt the natural units in which $\hbar=c=1$, $4\pi\epsilon_0=1$, and the electric and magnetic fields have squared-energy dimension. Thereby, the conversion of Volt/m and Tesla (T) to the natural system is as follows:  $1 \, \mbox{Volt/m}=2.27 \times 10^{-24} \, \mbox{GeV}^2$ and $1 \, \mbox{T} =  6.8 \times 10^{-16} \, \mbox{GeV}^2$, respectively. The metric convention is $\eta^{\mu\nu}=\mbox{diag}\left(+1,-1,-1,-1\right)$.
%

%
%
%%%%%%%%%%%%%%%%
\section{The non-linear axion-photon electrodynamics including the Carroll-Field-Jackiw term}
%%%%%%%%%%%%%%%%
\label{sec2}
We initiate with the description of the model whose Lagrangian density reads as follows :
\begin{eqnarray}\label{Lmodel}
{\cal L}=
%&=&
{\cal L}_{nl}({\cal F}_{0},{\cal G}_{0})
+\frac{1}{2} \, \left(\partial_{\mu}\phi \right)^{2}
-\frac{1}{2} \, m^2 \, \phi^2
+ g \, \phi \, {\cal G}_{0}
%\nonumber \\
%&&
+ \frac{1}{4} \, \epsilon^{\mu\nu\kappa\lambda} \, v_\mu \, A_{0\nu} \, F_{0\kappa\lambda}-J_{\mu}\,A_{0}^{\;\,\mu} \; ,
\end{eqnarray}
where $\mathcal{L}_{nl}({\cal F}_{0},{\cal G}_{0})$  denotes the most general Lagrangian of a non-linear electrodynamics that is function of the Lorentz- and gauge-invariant bilinears : $ {\cal F}_{0}=-\frac{1}{4} \, F_{0\mu\nu}^{2}=\frac{1}{2} \, \left( {\bf E}_{0}^2-{\bf B}_{0}^2\right)$  and ${\cal G}_{0}=-\frac{1}{4} \, F_{0\mu\nu}\widetilde{F}_{0}^{\;\,\,\mu\nu}={\bf E}_{0}\cdot{\bf B}_{0}$.
These definitions introduce the antisymmetric field strength tensor as
$F_{0}^{\;\,\,\mu\nu}=\partial^{\mu}A_{0}^{\;\,\nu}-\partial^{\nu}A_{0}^{\;\,\mu}=\left( \, -E_{0}^{\;\,i} \, , \, -\epsilon^{ijk}B_{0}^{\;\,k} \, \right)$,
and the correspondent dual tensor is $\widetilde{F}_{0}^{\;\,\,\mu\nu}=\epsilon^{\mu\nu\alpha\beta}F_{0\alpha\beta}/2=\left( \, -B_{0}^{\;\,i} \, , \, \epsilon^{ijk}E_{0}^{\;\,k} \, \right)$, which satisfies the Bianchi identity $\partial^{\mu}\widetilde{F}_{0\mu\nu}=0$. The CFJ term introduces the background $4$-vector $v^{\mu}=(v^{0},{\bf v})$ whose components do not depend on the space-time coordinates. It has mass dimension in natural units and is responsible for the Lorentz symmetry violation in the gauge sector of the model.
In addition, $\phi$ is the axion scalar field with mass $m$, and $g$ is the non-minimal
coupling constant (with length dimension) of the axion with the electromagnetic field, {\it i.e.}, the usual coupling with the ${\cal G}_{0}$-invariant in the axion-photon model. There are many investigations and experiments to constraint the possible regions in the space of the parameters $g$ and $m$, which still remains with a wide range of values, depending on the phenomenological scale in analysis.
We expand the abelian gauge field as $A_{0}^{\;\,\mu}=a^{\mu}+A_{B}^{\;\;\,\,\mu}$, in which $a^{\mu}$ is the photon $4$-potential,
and $A_{B}^{\;\;\,\,\mu}$  denotes a background potential. In this conjecture, the tensor $F_{0}^{\;\,\,\mu\nu}$ is also written as the
combination $F_{0}^{\;\,\,\mu\nu}=f^{\mu\nu} \, + \, F_{B}^{\;\;\,\mu\nu}$, in which $f^{\mu\nu}=\partial^{\mu}a^{\nu}-\partial^{\nu}a^{\mu}=\left( \, -e^{i} \, , \, -\epsilon^{ijk}b^{k} \, \right)$ is the EM field strength tensor that propagates in the space-time, and $F_{B}^{\;\;\,\mu\nu}=\left( \, -E^{i} \, , \, -\epsilon^{ijk}B^{k} \, \right)$ corresponds to the EM background field. The notation of the $4$-vector and tensors with index
$(B)$ indicates that it is associated with the background. At this stage, we consider the general case in
which the background depends on the space-time coordinates. Under this prescription, we also expand the Lagrangian (\ref{Lmodel})
around the background up to second order in the propagating field $a^{\mu}$ to yield the expression
\begin{eqnarray} \label{L4}
{\cal L}^{(2)}  \!&=&\!  -\frac{1}{4} \, c_{1} \, f_{\mu\nu}^{\, 2}
-\frac{1}{4} \, c_{2} \, f_{\mu\nu}\widetilde{f}^{\mu\nu}
+\frac{1}{8} \, Q_{B\mu\nu\kappa\lambda} \, f^{\mu\nu}f^{\kappa\lambda}
\nonumber \\
&&
\hspace{-0.5cm}
+\frac{1}{2} \, \left(\partial_{\mu}{\tilde\phi}\right)^{2}
-\frac{1}{2} \, m^2 \, \tilde{\phi}^2 - \frac{1}{2} \, g \, \tilde{\phi} \, \widetilde{F}_{B\mu\nu} \, f^{\mu\nu}
%\nonumber \\
%&&
%\hspace{-0.5cm}
+ \frac{1}{4} \, \epsilon^{\mu\nu\kappa\lambda} \, v_\mu \, a_{\nu} \, f_{\kappa\lambda} -\Bar{J}_{\nu}\,a^{\nu} \; ,
\end{eqnarray}
where $\Bar{J}_{\nu}= J_\nu -\partial^\mu\,(H_{B\mu\nu}) - v^\mu \widetilde{F}_{B\mu \nu} $ represent an effective external current that couples to the photon field; it includes an eventual matter current and the contributions that stem from the background electromagnetic fields. The tensors associated with this electromagnetic background are defined in what follows:
\begin{subequations}
\begin{eqnarray}
H_{B\mu\nu} \!&=&\! c_{1} \, F_{B\mu\nu} + c_{2} \, \widetilde{F}_{B\mu\nu}
+ \frac{g^2}{m^2} \, \mathcal{G}_B  \, \widetilde{F}_{B \mu\nu} \, ,
\\
Q_{B\mu\nu\kappa\lambda} \!&=&\! d_{1} \, F_{B \mu\nu} \, F_{B\kappa\lambda}
+d_{2} \, \widetilde{F}_{B \mu\nu} \, \widetilde{F}_{B \kappa\lambda}
%\nonumber \\
%&&
%\hspace{-0.5cm}
+d_{3} \, F_{B \mu\nu} \, \widetilde{F}_{B \kappa\lambda}
+ d_{3} \, \widetilde{F}_{B \mu\nu} \, F_{B \kappa\lambda} \; .
\end{eqnarray}
\end{subequations}
The axion field is shifted as $\phi \to \widetilde{\phi} + \phi_0$ in order to eliminate the $g \, \phi \, \mathcal{G}_{B} $ term that would appear
in the Lagrangian (\ref{L4}). The coefficients $c_{1}$, $c_{2}$, $d_{1}$, $d_{2}$ and $d_{3}$ are evaluated at ${\bf E}$ and ${\bf B}$, as follows :
\begin{eqnarray}\label{coefficients}
%&&
c_{1} = \left. \frac{\partial{\cal L}_{nl}}{\partial{\cal F}_{0}} \right|_{{\bf E},{\bf B}} ,\;
\left. c_{2}=\frac{\partial{\cal L}_{nl}}{\partial{\cal G}_{0}}\right|_{{\bf E},{\bf B}} , \,
\left. d_{1}=\frac{\partial^2{\cal L}_{nl}}{\partial{\cal F}_{0}^2}\right|_{{\bf E},{\bf B}} , \,
%\nonumber \\
%&&
\left. d_{2} = \frac{\partial^2{\cal L}_{nl}}{\partial{\cal G}_{0}^2}\right|_{{\bf E},{\bf B}}
, \, \left. d_{3}=\frac{\partial^2{\cal L}_{nl}}{\partial{\cal F}_{0}\partial{\cal G}_{0}}\right|_{{\bf E},{\bf B}} \, ,
\end{eqnarray}
that depend on the EM field magnitude and may also be functions of the space-time coordinates.
Following the previous definitions, the background tensors satisfy the properties $H_{B\mu\nu}=-H_{B\nu\mu}$, whereas $Q_{B\mu\nu\kappa\lambda}$ is symmetric under exchange $\mu\nu \leftrightarrow \kappa\lambda$, and antisymmetric under $\mu \leftrightarrow \nu$ and $\kappa \leftrightarrow \lambda$.  Note that the current $J^{\mu}$ couples to the external potential $A_{B}^{\;\,\,\,\mu}$, but this term and ${\cal L}_{nl}\left({\cal F}_{B},{\cal G}_{B}\right)$ are irrelevant for the field equations in which we are interested.
Using the minimal action principle by varying $a^{\mu}$, the Lagrangian \eqref{L4} yields the EM field
equations with source $\Bar{J}^{\mu}$
\begin{eqnarray}\label{EqGmunu}
%&&
\partial^\mu \left[ \, c_1 \, f_{\mu \nu} + c_2 \, \widetilde{f}_{\mu \nu} -
\frac{1}{2} \, Q_{B\mu \nu \kappa \lambda} \, f^{\kappa \lambda}\, \right]
%\nonumber \\
%&&
+ \, v^\mu \,\tilde{f}_{\mu\nu} =
- g \, ( \partial^\mu \widetilde{\phi} ) \,
\widetilde{F}_{B\mu \nu} + \Bar{J}_\nu \; ,
\;\;\;\;\;
\end{eqnarray}
and the Bianchi identity remains the same one for the photon field, namely, $\partial_{\mu}\widetilde{f}^{\mu\nu}=0$. The action principle in relation to $\widetilde{\phi}$ in (\ref{L4}) yields the axion field equation evaluated at the EM background :
 \begin{eqnarray}\label{eqescalar}
\left(\Box+m^2\right)\widetilde{\phi}=
-\frac{1}{2} \, g \, \widetilde{F}_{B\mu\nu} \, f^{\mu\nu}
\; .
\end{eqnarray}

Since we consider a uniform magnetic background field, the $c_2$- and $d_3$-coefficients of the expansion vanish for most examples of non-linear EDs known in the literature, such as Euler-Heisenberg, Born-Infeld, ModMax, Logarithmic and some others, where the corresponding non-linear Lagrangian densities depend on the square of the ${\cal G}$-invariant. These considerations simplify the results that we shall work out in the next Sections ahead. The usual axionic ED coupled to the CFJ-term is recovered whenever $d_{1} \rightarrow 0$, $d_{2} \rightarrow 0$ and $c_{1} \rightarrow 1$ in all the cases of non-linear ED mentioned previously.
%

%{\color{red} Since we consider a uniform magnetic background field, the $c_2$- and $d_3$-coefficients of the expansion vanish for all the examples of non-linear EDs known in the literature, as Euler-Heisenberg, Born-Infeld, ModMax, Logarithmic and others, where the correspondent non-linear lagrangian depend on the ${\cal G}$-invariant elevated to the squared. Therefore, the CPT invariance is restored in the linearized theory in all these cases, so we can consider $c_2=0$ and $d_3=0$ in the previous expressions. These considerations simplify the results that we will obtain in the next sections ahead.}

%
\section{The dispersion relations in presence of a uniform magnetic field}
\label{sec3}
In this Section, we obtain the dispersion relations of the axion and photon fields in a uniform magnetic background. Thus, we can take ${\bf E}= {\bf 0}$ in the equations of the section \ref{sec2}. Thereby, from now on, all the coefficients defined in (\ref{coefficients}) are not space-time dependent; they actually depend only on the magnetic vector ${\bf B}$. We start the description of the field propagating with the equations written in terms of ${\bf e}$ and ${\bf b}$, in the presence of constant and uniform magnetic background field. For the analysis of the free wave propagation, we just consider the linear terms in ${\bf e}$, ${\bf b}$ and $\widetilde{\phi}$, as well as, the equations with no source, ${\bf \Bar{J}}= \bf{0}$ and $\Bar{\rho}=0$. Under these conditions, the electrodynamics equations in terms of the propagating vector field are read below :
\begin{subequations}\label{subeqs}
\begin{eqnarray}
\nabla\cdot{\bf D} \!&=&\! {\bf v}\cdot {\bf b} \; ,
\\ %\;\;\; , \;\;\;
\nabla\times{\bf e}+\frac{\partial {\bf b}}{\partial t} \!&=&\! {\bf 0} \; ,
\label{divDrote}
\\
\nabla\cdot{\bf b} \!&=&\! 0 \;,
\\  %\;\;\;
\nabla\times{\bf H} + {\bf v}\times{\bf e} \!&=&\! v^0\,{\bf b} + \frac{\partial {\bf D}}{\partial t}\; ,
\label{divbrotH}
\end{eqnarray}
\end{subequations}
where the vectors ${\bf D}$ and ${\bf H}$ are, respectively, given by
\begin{subequations}
\begin{eqnarray}
{\bf D} \!&=&\! c_{1} \, {\bf e}
%+ c_{1} \, {\bf E}
+d_{2} \, {\bf B} \, ({\bf B}\cdot{\bf e}) + g \, \widetilde{\phi} \, {\bf B} \, , \;\;\;\;
\label{D}
\\
{\bf H} \!&=&\! c_{1} \, {\bf b}
%+ c_{1} \, {\bf B}
- d_{1} \, {\bf B} \, ({\bf B}\cdot{\bf b}) \, . \;\;\;\;
\label{H}
\end{eqnarray}
\end{subequations}
The scalar field equation (\ref{eqescalar}) in terms of the magnetic background field leads to
\begin{eqnarray}\label{EqscalarEB}
\left(\Box+m^2\right)\widetilde{\phi} = g \left({\bf e}\cdot{\bf B}\right) \; .
\end{eqnarray}
We substitute the plane wave solutions of ${\bf e}$, ${\bf b}$ and $\widetilde{\phi}$ in the field equations \eqref{subeqs}(a-d)
and \eqref{EqscalarEB}. Eliminating conveniently the amplitudes of ${\bf b}$ and $\widetilde{\phi}$ in terms of the electric field amplitude,
the wave equation in the momentum space is read below :
\begin{eqnarray}\label{Mijej}
M^{ij}(\omega,{\bf k}) \, e_{0}^{\,\,j}= 0 \; ,
%-\frac{g^2}{c_1} \, \frac{ \omega \, B_{i}-({\bf k}\times{\bf E})_{i} }{{\bf k}^2-\omega^2+m^2} \, \omega \, ({\bf E}\cdot{\bf B}) (2\pi)^{4}
%\delta^{3}({\bf k}) \, \delta(\omega) \; ,
\end{eqnarray}
where $e_{0}^{\,\,j} \, (j=1,2,3)$ are the components of the electric amplitude, and the matrix elements $M^{ij}$ are given by
\begin{equation}\label{Mij}
M^{ij}(\omega,{\bf k}) = a \, \delta^{ij} + b \, k^i \, k^j
+ c\, B^i \, B^j
%+
%\nonumber \\
%&&
%\hspace{-0.5cm}
+ d \left({\bf B\cdot k}\right) \left(B^i \, k^j + B^j \, k^i\right)
%\nonumber \\
%&&
%\hspace{-0.5cm}
- i\, \epsilon^{ijm} \, \left(\,v^0\,k^m \, - \, \omega \, v^m\,\right) \; ,
\end{equation}
whose the coefficients $a$, $b$, $c$ are defined by
\begin{subequations}
\begin{eqnarray}
a \!&=&\! \omega^2-{\bf k}^2+d \, \left({\bf k} \times {\bf B}\right)^2
\; , \;
\label{a}
\\
b \!&=&\! 1-d \, {\bf B}^2 \; ,
\label{b}
\\
c \!&=&\! \xi(\omega,{\bf k}) \, \omega^2   - d \, {\bf k}^2  \; ,
\label{c}
\\
\xi(\omega,{\bf k}) \!&=&\! f + \frac{g_{a}^2}{{\bf k}^2-\omega^2+m^2} \; ,
\label{xi}
\end{eqnarray}
\end{subequations}
in which $d:=d_{1}/c_{1}$, $f:=d_{2}/c_{1}$ and $g_{a}:=\sqrt{g^2/c_1}$ for simplicity in the equations. Thus, the non-linearity evaluated on the magnetic background is manifested in the parameters $d$ and $f$, and the coupling constant $g_{a}$ corrects the axion coupling constant with the coefficient $c_{1}$. Notice that the $b$-coefficient depends only on the magnetic background, but the others one depends on the $\omega$-frequency and on the ${\bf k}$-wave vector.

Back to the expressions of ${\bf D}$ and ${\bf H}$ in (\ref{D})-(\ref{H}) with the plane wave solutions,
the components of ${\bf D}$ and ${\bf H}$ in terms of the electric and magnetic amplitudes can be written as
\begin{eqnarray}
D_{i}=\epsilon_{ij}({\bf k},\omega)\,e_{j}
\hspace{0.3cm} \mbox{and} \hspace{0.3cm}
H_{i}=(\mu_{ij})^{-1} \, b_{j} \; ,
\end{eqnarray}
where $\epsilon_{ij}$ and $(\mu_{ij})^{-1}$ are the permittivity and permeability (inverse) tensors,
respectively,
\begin{subequations}
\begin{eqnarray}
\epsilon_{ij}({\bf k},\omega) \!&=&\! c_1 \, \delta_{ij} + c_{1} \, \xi({\bf k},\omega) \, B_{i} \, B_{j} \; ,
\label{epsilon}
\\
(\mu_{ij})^{-1} \!&=&\! c_{1} \, \delta_{ij}-d_{1} \, B_{i}\,B_{j} \; .
\label{mu}
\end{eqnarray}
\end{subequations}
The permeability tensor is obtained by computing the inverse of (\ref{mu})
\begin{eqnarray}\label{muij}
\mu_{ij}=\frac{1}{c_1} \frac{ \left(1-d\,{\bf B}^2\right)\delta_{ij}+d\, B_{i}\,B_{j} }{1-d\,{\bf B}^2} \; .
\end{eqnarray}
Notice that the electric permittivity depends on the $\omega$-frequency and the
${\bf k}$-wave vector due to the axion coupling $g \neq 0$. Also, the definition of these tensors
do not include the components of the CFJ $4$-vector $v^{\mu}$. Thereby, this LSV scenario does not
contribute with the physical properties of the tensors.

According to the works of refs. \cite{Adam,Baeta1,Baeta2}, the $v^0$-component may induce contributions that violate the causality and stability. For this reason, we shall adopt a space-like CFJ 4-vector, {\it i.e.}, $v^0=0$ in the matrix element $M^{ij}$ from eq. (\ref{Mij}). The dispersion relations come from the non-trivial solutions to the wave equation (\ref{Mij}). The condition for non-trivial solutions is $\det M^{ij} = 0$; for the space-like case of the CFJ background, it is reduced to an $\omega$-polynomial equation:

%
%{\color{red} For reasons of preservation of the unitarity and causality in the model \cite{Adam,Baeta1,Baeta2}}, we choose a space-like for the CFJ $4$-vector, {\it i.e.}, $v^0=0$ in the matrix element $M^{ij}$ from (\ref{Mij}).
%The dispersion relations come from the non-trivial solutions of the wave equation (\ref{Mij}). The condition for the non-trivial solution is $\det M^{ij} = 0$, that for the space-like case in CFJ, is reduced to $\omega$-polynomial equation:
%
%
%{\color{red}
%
%\begin{widetext}
\begin{eqnarray}\label{eq}
a^3+a^2 \left[ b\,{\bf k}^2+2d\,({\bf B}\cdot{\bf k})^2 \, \right]
+ac\left[ a\,{\bf B}^2+b({\bf B}\times{\bf k})^2\right]
-a\,d^2\,({\bf B}\cdot{\bf k})^{2}\,({\bf B}\times{\bf k})^2
\nonumber \\
-\left[a{\bf v}^2+c({\bf B}\cdot{\bf v})^2\right]\omega^2
%\nonumber \\
-({\bf v}\cdot{\bf k})
\left[\,b({\bf v}\cdot{\bf k})+2d\left({\bf B}\cdot{\bf k}\right)\left({\bf B}\cdot{\bf v}\right) \, \right]\omega^2=0 \; .
\;\;\;\;
\end{eqnarray}
%\end{widetext}
%
%
Back to eqs. (\ref{a})-(\ref{xi}), notice that the coefficient a takes into account non-linearity by means of the piece $d \, ({\bf B} \times {\bf k})^2 = d \, [\,{\bf B}^2\, {\bf k}^2 – ({\bf B}\cdot{\bf k})^2\, ]$,
$b$ incorporates non-linearity in the piece $d\,{\bf B}^2$ . The coefficient $c$, on the other hand, splits into non-linearity (the $d$- and $f$-terms) and axionic (the axion mass, $m$, and the coupling constant, $g_a$ ) effects: $c = (f \, \omega^2 – d\,{\bf k}^2 ) – g_a^2\, \omega^2 /( \omega^2 – {\bf k}^2 – m^2)$.
Finally, the LSV appears represented by the background vector, ${\bf v}$. It is worthy to remark that no term in the dispersion relation (\ref{eq}) couples the three effects together. We actually mean that
non-linearity, axionic and LSV effects interfere with one another only in pairs. No
term is present in which parameters of the three different effects appear grouped together in
a product. However, below, in discussing the effective masses of axion-photon coupled system, it will become clear that three effects mix up to give these effective masses.

We would like to point out that, if only non-linearity is considered, the dispersion relations do not exhibit momentum dependence, so no dispersive effect shows up. By introducing the axion sector, the profile changes and dispersive effects emerges. Now, if LSV is also considered in addition to non-linearity and axionic physics, not only dispersion is enforced, but another fact should be highlighted: photonic dispersion relations show that the modulus of the LSV background vector, $|{\bf v}|=v$, along with the axion mass endows the photon with an effective mass, $m_{eff}$. For instance, let us consider the rest frame ($k=0$ with the identification $\omega^2 = m_{eff}^2$), as well as ${\bf B} = B \, \hat{{\bf z}} $ and ${\bf v} = v \, \hat{{\bf y}}$ in eq. \eqref{eq}. Therefore, it is possible to show that the effective masses for the photon and the axion correspond to the roots of the equation
\begin{eqnarray}\label{eq_meff}
(1 + f \, B^2) \, m_{eff}^4 - \left[ \, (1 + f \, B^2)\, m^2  + \, g_a^{2} \, B^2 +  \frac{v^2}{c_1^2} \, \right] \, m_{eff}^2  + \frac{m^2\,v^2}{c_1^2} = 0 \; .
\end{eqnarray}
The solutions of this quartic equation are
\begin{subequations}
\begin{eqnarray}
m_{eff(1)}^2 &=& \frac{(1+f\,B^2)\,m^2+g_a^2\,B^2+v^2/c_1^2}{2 \left(1+f\,B^2\right)}
\nonumber \\
&&
\hspace{-0.5cm}
-\frac{\sqrt{\left((1+f\,B^2)m^2+g_a^2\,B^2+v^2/c_1^2\right)^2-4 m^2v^2\left(1+f\,B^2\right)/c_1^2}}{2 \left(1+f\,B^2\right)} \; ,
\label{mgamma1}
\\
m_{eff(2)}^2 &=& \frac{(1+f\,B^2)\,m^2+g_a^2\,B^2+v^2/c_1^2}{2 \left(1+f\,B^2\right)}
\nonumber \\
&&
\hspace{-0.5cm}
+\frac{\sqrt{\left((1+f\,B^2)m^2+g_a^2\,B^2+v^2/c_1^2\right)^2-4 m^2v^2\left(1+f\,B^2\right)/c_1^2}}{2 \left(1+f\,B^2\right)} \; .
\label{mgamma2}
\end{eqnarray}
\end{subequations}
In the limit $g_{a}\rightarrow 0$, the first solution (\ref{mgamma1}) is reduced to
\begin{eqnarray} \label{m_eff_1}
m_{eff(1)}=\frac{v}{\sqrt{c_1(c_1+d_2\,B^{2})}} \; ,
\end{eqnarray}
whereas, when $v \rightarrow 0$, the effective mass is null. Consequently, in the uncoupled limit ($g_a\rightarrow 0$),
the CFJ parameter $(v)$ gives an effective mass with the correction of the non-linearity evaluated at the magnetic background
field. Considering the same limit of $v \rightarrow 0$, the second solution (\ref{mgamma2}) is reduced to the expression
\begin{eqnarray} \label{m_eff_2}
m_{eff(2)}=\sqrt{m^2 + \frac{g^2\,B^2}{c_1(c_1+d_2\,B^2)} } \; ,
\end{eqnarray}
where the axion mass, and the coupling constant have a fundamental role for the effective mass. Thus, the axion mass is corrected by the coupling constant and by the magnetic field. In the uncoupled limit, $g_{a}\rightarrow 0$,  the second root
(\ref{mgamma2}) yields the axion mass, {\it i.e.}, $m_{eff(2)}=m$.

Now let us return to eq. (\ref{eq}). The general solution is quite involved in view of the coefficients (\ref{a})-(\ref{xi}). For simplicity, we consider the two cases below :
%

%We back to the solution to the equation (\ref{eq}) is hard to be worked out in view of the coefficients (\ref{a})-(\ref{xi}). For simplicity, we consider the two cases below :
%

%
%
\begin{enumerate}[label=(\alph*)]
\item The case of the vectors ${\bf B}$, ${\bf k}$ and ${\bf v}$ perpendiculars among themselves : ${\bf B}\cdot{\bf k}={\bf B}\cdot{\bf v}={\bf k}\cdot{\bf v}=0$. Considering this condition, the equation (\ref{eq}) is reduced to :
\begin{equation}\label{eqperp}
\omega_{\perp}^2\left[\omega_{\perp}^2-k^2+d\,B^{2}\,k^2\right]\left[\left(1+\xi \, B^2\right)\omega_{\perp}^2-k^2-v^2\right]=0 \; ,
\end{equation}
where we denote the perpendicular frequency $\omega_{\perp}$, and $B$, $k$ and $v$ are the magnitudes of the previous vectors.
The first solution is $\omega_{\perp}=0$, and the non-trivial solutions from (\ref{eqperp}) are given by
\begin{widetext}
\begin{subequations} \label{omegasperp}
\begin{eqnarray}
\omega_{1\perp}(k) \!&=&\! k \, \sqrt{1-d \, B^2 } \; ,
\label{omega1perp} \\
\omega_{2\perp}(k) \!&=&\! \left\{ \, \frac{ k^2+v^2+g_a^2\,B^2+(1+f \, B^2)(k^2+m^2)}{2 \, (1+f \, B^2)}-
\right.
\nonumber \\
&&
\hspace{-0.5cm}
\left.
-\frac{\sqrt{\left( k^2+v^2+g_a^2\,B^2+(1+f \, B^2)(k^2+m^2) \right)^2-4 \left(1+f\,B^2\right) \left(k^2+m^2\right) \left(k^2+v^2\right)}}{2 \, (1+f\,B^2)} \, \right\}^{1/2} ,
\label{omega2perp}
\nonumber \\
\\
\omega_{3\perp}(k) \!&=&\! \left\{ \, \frac{k^2+v^2+g_a^2\,B^2+(1+f \, B^2)(k^2+m^2)}{2 \, (1+f \, B^2)}+
\right.
\nonumber \\
&&
\hspace{-0.5cm}
\left.
+\frac{\sqrt{\left( k^2+v^2+g_a^2\,B^2+(1+f \, B^2)(k^2+m^2) \right)^2-4 \left(1+f\,B^2\right) \left(k^2+m^2\right) \left(k^2+v^2\right)}}{2 \, (1+f\,B^2)} \, \right\}^{1/2} . \label{omega3perp}
\nonumber \\
\end{eqnarray}
\end{subequations}
\end{widetext}
The analysis of the limits to establish comparisons with the results in the literature is immediate. The limits $f \rightarrow 0$ and $c_1 = 1$ yield the dispersion relations of the axionic ED coupled to CFJ term in the presence of an external magnetic field. Furthermore, if we also take $g_a \rightarrow 0$, the dispersion relations are reduce to $\omega_{2\perp}(k)=\sqrt{(k^2+v^2)(1+f\,B^2)^{-1}}$ and $\omega_{3\perp}(k)=\sqrt{k^2+m^2}$ for $m > v$. Note that in $\omega_{2\perp}(k)$, occurs the characteristic effect of CFJ, where the Lorentz-breaking parameter gives a small mass for the photon.
These results confirm the roots of the eq. (\ref{eq_meff}) in the rest frame ($k=0$).  % and $\omega_{2\perp}(k)=\sqrt{k^2+m^2}$ and $\omega_{3\perp}(k)=\sqrt{k^2+v^2}$ if $v>m$.
The usual Maxwell limit reduces all the frequencies to : $\omega_{1\perp}(k)=\omega_{2\perp}(k)=k$
and $\omega_{3\perp}(k)=\sqrt{k^2+m^2}$. The refractive (perpendicular) index are defined by
\begin{eqnarray}\label{nperp}
n_{i\perp}({\bf k})=\frac{|{\bf k}|}{\omega_{i\perp}({\bf k})} \; , \; (i=1,2,3) \; .
\end{eqnarray}
\item The second case consists in considering ${\bf v}$ orthogonal to both ${\bf B}$ and ${\bf k}$, but ${\bf B}$ parallel to ${\bf k}$ :
${\bf B}\cdot{\bf v}={\bf k}\cdot{\bf v}=0$ and ${\bf B}\cdot{\bf k}=B\,k$. In this case, the equation (\ref{eq}) is
\begin{equation}\label{eqpara}
\omega_{\parallel}^2 \, \left(\omega_{\parallel}^2-k^2\right)\left[ \, \left(1+\xi \, B^2\right)\left(\omega_{\parallel}^2-k^2\right)-v^2 \, \right]=0
\, ,
\end{equation}
where $\omega_{\parallel}$ is now the frequency for ${\bf B}$ parallel to ${\bf k}$. The trivial solution is $\omega_{\parallel}=0$,
and the others solutions are read below :
\begin{widetext}
\begin{subequations} \label{omegasparal}
\begin{eqnarray}
\omega_{1\parallel}(k) \!&=&\! k \; ,
\label{omega1par}
\\
\omega_{2\parallel}(k) \!&=&\! \left\{ \, \frac{ \left(2 k^2+m^2\right)\left(1+f\,B^2\right)+v^2+ g_a^2\,B^2}{2\,(1+f\,B^2)}
\right.
\nonumber \\
&&
\left.
-\frac{\sqrt{\left[ \, g_a^2\,B^2+m^2(1+f\,B^2) \, \right]^2+2 v^2 \, g_a^2\,B^2-2v^2\,m^2 \left(1+f\,B^2\right)+v^4}}{2 \, (1+f \, B^2)} \,\right\}^{1/2} \; ,
\label{omega2par}
\nonumber \\
\\
\omega_{3\parallel}(k) \!&=&\! \left\{ \, \frac{ \left(2 k^2+m^2\right)\left(1+f\,B^2\right)+v^2+ g_a^2\,B^2}{2\,(1+f\,B^2)}
\right.
\nonumber \\
&&
\left.
+\frac{\sqrt{\left[ \, g_a^2\,B^2+m^2(1+f\,B^2) \, \right]^2+2 v^2 \, g_a^2\,B^2-2v^2\,m^2 \left(1+f\,B^2\right)+v^4}}{2 \, (1+f \, B^2)} \,\right\}^{1/2} \nonumber \\
\; .
\label{omega3par}
\end{eqnarray}
\end{subequations}
\end{widetext}
The first solution (\ref{omega1par}) is the usual photon DR due to ${\bf B}\times{\bf k}={\bf 0}$ in the $a$-parameter in (\ref{a}).
The limits of $f \rightarrow 0$ and $c_1=1$ also recover the DRs of the axionic ED coupled to CFJ term in the presence of the external magnetic field $B$.
The limit $g_a \rightarrow 0$, when the axion is decoupled from the CFJ ED, the DRs are reduced to the results :
$\omega_{2\parallel}(k)=\sqrt{k^2+v^2\,(1+f\,B^2)^{-1}}$ and $\omega_{3\parallel}(k)=\sqrt{k^2+m^2}$ for $m > v$. %, and $\omega_{2\perp}(k)=\sqrt{k^2+m^2}$ and $\omega_{3\perp}(k)=\sqrt{k^2+v^2}$, if $v>m$.
This confirm the same results recovered in the case (a). The correspondent refractive (parallel) index are defined by
\begin{eqnarray}\label{npara}
n_{i\parallel}({\bf k})=\frac{|{\bf k}|}{\omega_{i\parallel}({\bf k})} \; , \;  (i=1,2,3) \; .
\end{eqnarray}
where we must substitute the DRs (\ref{omega1par})-(\ref{omega3par}). Notice that, in both the cases (a) and (b), the refractive index of
the medium depends on the modulus ${\bf k}$, so, consequently, it depends on the wavelength, as
$\lambda=2\pi/|{\bf k}|$.
%

%{\color{red}
%
To close this section, in possess of the set of dispersion relations (\ref{omegasperp}) and (\ref{omegasparal}), we call back one of the motivations to do this work, namely, to keep track of how the three different physical scenarios we bring together in the action of eq. (\ref{Lmodel}) interfere with one another, which is manifested by means of the terms coupling the parameters of the different scenarios. Keeping in mind that the coefficients $c_1$ , $f$ and $d$ express the non-linearity and that $g_a$ incorporates the axion-photon coupling and the coefficient $c_1$, the presence of the denominator $(1 + f B^2)$, common to all frequency solutions, in combination with the terms in $m^2$, $v^2$, $g_a B^2$, $f B^2 m^2$ and $f B^2 m^2 v^2$ , as it appears in eqs. (\ref{omega1perp})-(\ref{omega3perp}) and (\ref{omega1par})-(\ref{omega3par}), shows in an explicit way how the three different physics mix among themselves to produce tiny effects in optical quantities like phase and group velocities and refraction indices.
%}
%{\color{blue}
The explicit forms of the coefficients in terms of the non-linear electrodynamic models of Euler-Heisenberg, Born-Infeld and ModMax will be shown in the next section, namely, by equations (\ref{cEH}),(\ref{cBI}) and (\ref{coefficientsMM}), respectively.

%}

%
\end{enumerate}
\section{The birefringence phenomenon}
\label{sec4}
%

%
%The birefringence phenomenon is manifested by the difference between the parallel and perpendicular refractive indices:
%

Birefringence is an optical property of an anisotropic medium expressed by the dependence of the refractive index on the polarization and direction of propagation of an electromagnetic wave. Just to recall, the polarization conventionally refers to the configuration of the electric field of the wave. However, in the previous Section, we have worked out refraction indices associated to the propagation of the waves in two situations: perpendicular and parallel to the background magnetic field: ${\bf k}\cdot{\bf B} = 0$  and  ${\bf k}\cdot{\bf B} = |{\bf k}| |{\bf B}|$, respectively, with no reference to the polarization  established by the electric field. Eqs. (\ref{nperp}) and (\ref{npara}) explicitly show how the non-linearity – manifested by the external magnetic field – the axion parameters and the LSV vector interfere with one another in the expressions for the perpendicular and parallel refraction indices. And we would like to stress that we are here adopting the point of view that the phenomenon of birefringence manifests itself by the difference between the refractive indices of eqs. (\ref{nperp}) and (\ref{npara}) , as defined below,

\begin{eqnarray}\label{Deltani}
\Delta n_{ij} ({\bf k}) = n_{i\parallel}({\bf k}) - n_{j\perp}({\bf k}) \; , \; (i,j=1,2,3) \; ,
\end{eqnarray}
 where we are contemplating the cases in which $i = j$  and $i \neq j$; in general, $\Delta n_{ij} \neq 0$, and it depends on the wavelength, which characterizes dispersive propagation. Notice also that  $\Delta n_{ij} \neq \Delta n_{ji}$ according to the definition (\ref{Deltani}). The difference between the refraction indices in these situations is exclusively due to the choice of the wave propagation direction with respect to the external {\bf B}-field.
Substituting the results from the previous section, the variation of refractive index in the case of $i=j$ are read
\begin{subequations}
\begin{eqnarray}
\Delta n_{11} \!&=&\! 1-\frac{1}{\sqrt{1-d\,B^2}}  \; ,
\label{Deltan1}
\\
\Delta n_{22} (k) \!&\simeq&\!
\sqrt{ \frac{1+f\,B^2}{1+f\,B^2+v^2/k^2} }
-\sqrt{ \frac{1+f\,B^2}{1+v^2/k^2} }\left[ 1+\frac{g_a^2\,B^2/2}{m^2+f\,B^2(k^2+m^2)-v^2} \right]  \; ,
\label{Deltan2}
\\
\Delta n_{33} (k) \!&\simeq&\!
\frac{1}{(1+m^2/k^2)^{3/2}} \frac{m^2-v^2}{(1+f\,B^2)m^2-v^2}
%\nonumber \\
%&&
%\hspace{-0.5cm}
%\times \;
\frac{g_a^2\,B^2/2}{m^2+f\,B^2(k^2+m^2)-v^2} \; ,
%
%\frac{g_a^2\,B^2}{(k^2+m^2)^{3/2}} \frac{k^2}{(1+f\,B^2)\,m^2-v^2}
%\nonumber \\
%&&
%\hspace{-0.5cm}
%\times \;
%\frac{m^2-v^2}{m^2-v^2+f\,B^2(k^2+m^2)} \; ,
\label{Deltan3}
\end{eqnarray}
\end{subequations}
where we have considered that $g_a$ is very weak in comparison with the squared inverse of the magnetic background $(g_a^2 \, B\ll 1)$.
%, and the wave number $(k=2\pi/\lambda)$ is much greater than the CFJ parameter, {\it i. e.}, $k \gg v$.
The birefringence effects for $i \neq j$ are read below :
\begin{subequations}
\begin{eqnarray}
\Delta n_{12}(k) \!&\simeq&\! 1-\sqrt{\frac{1+f\,B^2}{1+v^2/k^2}}
\left[ 1+\frac{g_a^2\,B^2/2}{m^2+f\,B^2(k^2+m^2)-v^2} \right] \; ,
\label{Deltan12}
\\
\Delta n_{21}(k) \!&\simeq&\! \sqrt{ \frac{1+f\,B^2}{1+f\,B^2+v^2/k^2} } -\frac{1}{\sqrt{1-d\,B^2}}
%+\frac{k}{ \sqrt{k^2+\frac{v^2}{1+f\,B^2}}}
\; ,
\label{Deltan21}
\\
\Delta n_{13} (k) \!&\simeq&\! 1 - \frac{1}{\sqrt{1+m^2/k^2}}\left[1-
\frac{g_a^2\,B^2/2}{m^2+f \, B^2 \left(k^2+m^2\right)-v^2}\right]
%\nonumber \\
%&&
%\hspace{-0.5cm}
%+\frac{v^2}{2k^2} \left(\sqrt{1+f\,B^2}-\frac{1}{1+f\,B^2}\right)
\; ,
\label{Deltan13}
\\
\Delta n_{31} (k) \!&\simeq&\! \frac{1}{\sqrt{1 + m^2/k^2}}-\frac{1}{\sqrt{1-d\,B^2}}
%\nonumber \\
%&&
%\hspace{-1.5cm}
%+\frac{k}{\sqrt{k^2 + m^2}}
-\frac{m^2\,k}{(k^2+m^2)^{3/2}}\frac{g_a^2\,B^2/2}{m^2(1+f \, B^2)-v^2}
%\nonumber \\
%&&
%\hspace{-0.5cm}
%+\frac{v^2}{2k^2} \left(\sqrt{1+f\,B^2}-\frac{1}{1+f\,B^2}\right)
\; ,
\label{Deltan31}
\\
\Delta n_{23} (k) \!&\simeq&\! \sqrt{ \frac{1+f\,B^2}{1+f\,B^2+v^2/k^2} }-\frac{1}{\sqrt{1+m^2/k^2}}
\left[1-\frac{g_{a}^2\,B^2/2}{m^2+f\,B^2(k^2+m^2)-v^2}\right]
\; ,
\label{Deltan23}
\\
\Delta n_{32} (k) \!&\simeq&\!  \sqrt{ \frac{1+f\,B^2}{1+v^2/k^2} }
-\frac{1}{\sqrt{1+m^2/k^2}}
- \frac{g_a^2\,B^2/2}{[(1+f\,B^2)m^2-v^2]}\frac{m^2\,k}{(k^2+m^2)^{3/2}}
\nonumber \\
&&
\hspace{-0.5cm}
- \frac{g_a^2\,B^2/2}{m^2+f\,B^2(k^2+m^2)-v^2} \, \sqrt{\frac{1+f\,B^2}{1+v^2/k^2}}
\; .
\label{Deltan32}
\end{eqnarray}
\end{subequations}
%
%
%{\color{red}
Turning off the magnetic background $(B\rightarrow0)$, birefringence emerges  in all
the results with $\lim_{B\rightarrow 0}\Delta n_{ij} \neq 0$, for $i\neq j$, and $\lim_{B\rightarrow 0}\Delta n_{ij} = 0$ for $i=j$.
Therefore, in this limit of $B\rightarrow 0$, the birefringence phenomenon appears only due to the axion mass
and the CFJ parameter. Only the expression (\ref{Deltan1}) does not depend on the wavelength. For the usual Maxwell ED coupled to the axion and the CFJ term, the limits of $c_1 \rightarrow 1$, $d \rightarrow 0$ and $f \rightarrow 0$ yield the results below :
\begin{subequations}
\begin{eqnarray}
\Delta n_{11} \!&=&\! 0  \; ,
\label{Deltan1noED}
\\
\Delta n_{22}(k) \!&\simeq&\! -\frac{1}{\sqrt{ 1+v^2/k^2} }\frac{g^2\,B^2}{2\,(m^2-v^2)} \; ,
\label{Deltan2noED}
\\
\Delta n_{33}(k) \!&\simeq&\! \frac{1}{(1+m^2/k^2)^{3/2}} \frac{g^2\,B^2}{2\,(m^2-v^2)} \; ,
\label{Deltan3noED}
\\
\Delta n_{12}(k) \!&\simeq&\! 1-\frac{1}{\sqrt{1+v^2/k^2}}\left[1+\frac{g^2\,B^2}{2\,(m^2-v^2)}\right]\; ,
\label{Deltan21noED}
\\
\Delta n_{21}(k) \!&\simeq&\! \frac{1}{\sqrt{1+v^2/k^2}}-1  \; ,
\label{Deltan12noED}
\\
\Delta n_{13}(k) \!&\simeq&\! 1-\frac{1}{\sqrt{1+m^2/k^2}}\left[ 1-\frac{g^2\,B^2}{2\,(m^2-v^2)} \right] \; ,
\label{Deltan13noED}
\\
\Delta n_{31}(k) \!&\simeq&\! \frac{1}{\sqrt{1+m^2/k^2}}-1-\frac{g^2\,B^2}{2\,(m^2-v^2)}\frac{m^2\,k}{(k^2+m^2)^{3/2}}
\; ,
\label{Deltan13noED}
\\
\Delta n_{23}(k) \!&\simeq&\! \frac{1}{\sqrt{1+v^2/k^2}}-\frac{1}{\sqrt{1+m^2/k^2}} \left[ 1-\frac{g^2\,B^2}{2\,(m^2-v^2)} \right] \; ,
\label{Deltan23noED}
\\
\Delta n_{32}(k) \!&\simeq&\! \frac{1}{\sqrt{1+v^2/k^2}}-\frac{1}{\sqrt{1+m^2/k^2}}
\nonumber \\
&&
\hspace{-0.5cm}
- \frac{g^2\,B^2}{2\,(m^2-v^2)} \frac{m^2\,k}{(1+m^2/k^2)^{3/2}}
- \frac{g^2\,B^2}{2\,(m^2-v^2)}\frac{1}{\sqrt{1+v^2/k^2}} \; ,
\label{Deltan32noED}
\end{eqnarray}
\end{subequations}
where now the parameters $g$, $B$ and the $v$-CFJ have a fundamental rule for the birefringence phenomenon.
It is worth to highlight that if we consider a massless axion, the birefringence is null only in (\ref{Deltan13noED}).
The limit $g_a \rightarrow 0$, for which we have a non-linear ED coupled to the CFJ term, the results (\ref{Deltan1})-(\ref{Deltan32}) are reduced to
\begin{subequations}
\begin{eqnarray}
\Delta n_{11} \!&=&\! 1-\frac{1}{\sqrt{1-d\,B^2}}  \; ,
\label{Deltan1g0}
\\
\Delta n_{22} (k) \!&=&\! \sqrt{ \frac{1+f\,B^2}{1+f\,B^2+v^2/k^2} }-\sqrt{\frac{1+f\,B^2}{1+v^2/k^2}}  \; ,
\label{Deltan2g0}
\\
\Delta n_{33} (k) \!&=&\! 0 \; ,
\label{Deltan3g0}
\\
\Delta n_{12}(k) \!&=&\! 1-\sqrt{ \frac{1+f\,B^2}{1+v^2/k^2} }\; ,
\label{Deltan12g0}
\\
\Delta n_{21}(k) \!&=&\! \sqrt{ \frac{1+f\,B^2}{1+f\,B^2+v^2/k^2} } -\frac{1}{\sqrt{1-d\,B^2}} \; ,
\label{Deltan21g0}
\\
\Delta n_{13} (k) \!&=&\! 1 - \frac{1}{\sqrt{1 + m^2/k^2}}
\; ,
\label{Deltan13g0}
\\
\Delta n_{31} (k) \!&=&\! \frac{1}{\sqrt{1 + m^2/k^2}}-\frac{1}{\sqrt{1-d\,B^2}}
\; ,
\label{Deltan31g0}
\\
\Delta n_{23} (k) \!&=&\! \sqrt{ \frac{1+f\,B^2}{1+f\,B^2+v^2/k^2} }-\frac{1}{\sqrt{1+m^2/k^2}}
\; ,
\label{Deltan23g0}
\\
\Delta n_{32} (k) \!&=&\!  \sqrt{ \frac{1+f\,B^2}{1+v^2/k^2} }-\frac{1}{\sqrt{1+m^2/k^2}}
\; .
\label{Deltan32g0}
\end{eqnarray}
\end{subequations}
In this case, the non-linearity plays a key role in the birefringence phenomenon.
We shall discuss ahead birefringence by contemplating three non-linear electrodynamic models :
Euler-Heisenberg, Born-Infeld and ModMax ED.
\begin{enumerate}[label=(\alph*)]
\item The Euler-Heisenberg ED is described by the Lagrangian :
\begin{eqnarray}\label{LEHapprox}
{\cal L}_{EH}({\cal F},{\cal G}) = {\cal F}+\frac{2\alpha^2}{45m_{e}^4} \left(\, 4 \, {\cal F}^2\,+\,7 \, {\cal G}^2 \,\right) \; ,
\end{eqnarray}
where $\alpha=e^2=(137)^{-1}=0.00729$ is the fine structure constant, and $m_{e}=0.5 \, \mbox{MeV}$ is the electron mass.
Taking this Lagrangian and applying the expansion presented in Section (\ref{sec2}), the coefficients read as below:
%
%\begin{subequations}
\begin{eqnarray}\label{cEH}
d^{EH}
%=\frac{d_{1}^{EH}}{c_{1}^{EH}}
\simeq \frac{16\alpha^2}{45m_{e}^{4}}
\hspace{0.4cm} \mbox{and} \hspace{0.4cm}
f^{EH} \simeq \frac{28\alpha^2}{45m_{e}^{4}} \; ,
\end{eqnarray}
%\end{subequations}
%
for a weak magnetic field. Substituting these coefficients in (\ref{Deltan1})-(\ref{Deltan3}), we obtain
\begin{subequations}
\begin{eqnarray}
\Delta n_{11}^{(EH)} \!&\simeq&\! -\frac{8\alpha^2B^2}{45m_{e}^{4}}  \; ,
\label{Deltan1EH}
\\
\Delta n_{22}^{(EH)} \!&\simeq&\! -\frac{14\alpha^2B^2}{45m_{e}^{4}}\frac{1}{(1+v^2/k^2)^{3/2}}
+\frac{g^2\,B^2}{2\,(m^2-v^2)}\frac{1}{(1+m^2/k^2)^{3/2}} \; ,
\label{Deltan2EH}
\\
\Delta n_{33}^{(EH)} \!&\simeq&\!  \frac{g^2\,B^2}{2\,(m^2-v^2)} \frac{1}{(1+m^2/k^2)^{3/2}} \; ,
\label{Deltan3EH}
\\
\Delta n_{12}^{(EH)} \!&\simeq&\! 1-\frac{1}{\sqrt{1+v^2/k^2}}\left[ \, 1+\frac{14\alpha^2B^2}{45m_{e}^{4}}
+\frac{g^2B^2}{2\,(m^2-v^2)} \, \right]  \; ,
\label{Deltan12EH}
\\
\Delta n_{21}^{(EH)} \!&\simeq&\! \frac{1}{\sqrt{1+v^2/k^2}}-1 -\frac{8\alpha^2B^2}{45m_{e}^{4}}  \; ,
\label{Deltan21EH}
\\
\Delta n_{13}^{(EH)} \!&\simeq&\! 1-\frac{1}{\sqrt{1+m^2/k^2}}\left[1-\frac{g^2\,B^2}{2\,(m^2-v^2)} \right] \; ,
\label{Deltan13EH}
\\
\Delta n_{31}^{(EH)} \!&\simeq&\! \frac{1}{\sqrt{1+m^2/k^2}}-1-\frac{8\alpha^2B^2}{45m_{e}^{4}}-\frac{g^2\,B^2}{2\,(m^2-v^2)}\frac{m^2\,k}{(k^2+m^2)^{3/2}} \; ,
\label{Deltan31EH}
\\
\Delta n_{23}^{(EH)} \!&\simeq&\! \frac{1}{\sqrt{1+v^2/k^2}} - \frac{1}{\sqrt{1+m^2/k^2}}\left[1-\frac{g^2\,B^2}{2\,(m^2-v^2)}\right] \; ,
\label{Deltan23EH}
\\
\Delta n_{32}^{(EH)} \!&\simeq&\! \frac{1}{\sqrt{1+m^2/k^2}} - \frac{1}{\sqrt{1+v^2/k^2}}\left[1+\frac{14\alpha^2B^2}{45m_{e}^4}+\frac{g^2\,B^2}{2\,(m^2-v^2)} \right]
\nonumber \\
&&
\hspace{-0.5cm}
-\frac{g^2\,B^2}{2\,(m^2-v^2)}\frac{m^2\,k}{(k^2+m^2)^{3/2}}
\; ,
\label{Deltan32EH}
\end{eqnarray}
\end{subequations}
where we have neglected terms with $\alpha^2\,g^2$.
Using the parameters previously defined, the solution (\ref{Deltan1EH}) yields the numeric value
\begin{eqnarray}
\frac{|\Delta n_{11}^{(EH)}|}{B^2} \simeq \frac{8\alpha^2}{45m_{e}^{4}}=5.3\times 10^{-24} \, \mbox{T}^{-2} \; ,
\end{eqnarray}
that is of the same order of the result presented by the PVLAS-FE experiment for vacuum magnetic birefringence,
{\it i.e.}, $\Delta n_{PVLAS-FE}/B^2=(19\pm 27) \times 10^{-24}\,\mbox{T}^{-2}$ \cite{25years}.
The result (\ref{Deltan2EH}) contains the contribution of $g^2$, and also of the $v$-parameter, whereas the
variation $\Delta n_{33}^{(EH)}$ is proportional to $g^2\,B^2$ and it does not depend on $\alpha$.
%The solution $\Delta n_{13}^{(EH)}$ is finite in the limit of $B \rightarrow 0$.
Turning off the magnetic background, we obtain $\Delta n_{ii}^{(EH)}=0$. In this limit, the CFJ $v$-parameter contributes
to the birefringence in $\Delta n_{ij}^{(EH)} (i\neq j)$, as follows :
\begin{subequations}
\begin{eqnarray}
\lim_{B\rightarrow 0}
\Delta n_{12}^{(EH)} \!&=&\! -\Delta n_{21}^{(EH)}=
1-\frac{k}{\sqrt{k^2+v^2}} \; ,
\label{Deltan12EHB0}
%\simeq -\frac{v^2}{2k^2}
\\
\lim_{B\rightarrow 0}
\Delta n_{13}^{(EH)} \!&=&\! -\Delta n_{31}^{(EH)} = 1-\frac{k}{\sqrt{k^2+m^2}} \; ,
\label{Deltan13EHB0}
%\simeq \frac{m^2}{2k^2} \; ,
\\
\lim_{B\rightarrow 0}
\Delta n_{23}^{(EH)} \!&=&\! - \Delta n_{32}^{(EH)} = \frac{k}{\sqrt{k^2+v^2}} - \frac{k}{\sqrt{k^2+m^2}} \; .
\label{Deltan23EHB0}
\end{eqnarray}
\end{subequations}
%
%
%In the case of $\Delta n_{21}^{(EH)}$, the CFJ $v$-parameter contributes to the birefringence when $B \rightarrow 0$ :
%
%
%In this same limit, the variation $\Delta n_{31}^{(EH)}$ is $\Delta n_{31}^{(EH)} \simeq -m^2/(2k^2)$
%if $m > v$, and $\Delta n_{31}^{(EH)} \simeq -v^2/(2k^2)$ if $v>m$.
These conditions constrain the CFJ parameter to depend on the range of the axion mass. This is the case of a ultra-light axion (ULA),candidate to DM, with mass lower-bounded according to $m \gtrsim 10^{-22} \, \mbox{eV} \, (2\sigma)$ from non-linear clustering \cite{marsh}, to be compared with $v \lesssim 10^{-23}-10^{-25}$ GeV \cite{yuri}. Notice that the limit $B \rightarrow 0$ is equivalent to the cases in which the non-linearity is absent $(\alpha \simeq 0)$, and also when the axion coupling constant disappears$(g\rightarrow 0)$. Thus, the only parameters that remain are $v$ and the axion mass. The mass for the free axion is not a sufficient parameter to guarantee birefringence, whereas if we have only the $v$-parameter, the birefringence of the CFJ ED is recovered \cite{cfj}.
\item The Born-Infeld ED is governed by the Lagrangian :
\begin{eqnarray}\label{BIg}
{\cal L}_{BI}({\cal F},{\cal G})= \beta^2 \left[\, 1-\sqrt{ \, 1-\frac{2{\cal F}}{\beta^2}-\frac{{\cal G}^2}{\beta ^4} \, } \, \right] \; ,
\end{eqnarray}
where $\beta$ is the critical field of this model and has squared mass dimension.
The usual Maxwell ED is recovered whenever $\beta \rightarrow \infty$. This is the well motivated non-linear theory in which a point-like charge exhibits a finite electric field at the origin. A maximum electric field produces a finite self-energy for the electron that fixes the BI-parameter at $\beta=1.187 \times 10^{20} \, \mbox{V}/\mbox{m}$, that in MeV scale is $\sqrt{\beta}=16$ MeV \cite{BornInfeld}. As an example in accelerators, the ATLAS collaboration constraints the $\beta$-parameter in the stringent bound of $\sqrt{\beta} \gtrsim 100$ GeV through the light-by-light scattering in Pb-Pb collisions \cite{Ellis}. Since our analysis is associated with the propagation effects in the linearized BI ED, we consider the BI $\beta$-parameter in the low-energy scale, $\sqrt{\beta}=16$ MeV. We understand that, upon the inclusion of radiative corrections, the renormalization group equations must exhibit a running of the $\beta$-parameter with energy and the external magnetic field, since the latter is present in the propagators and interactions vertices. This then means that the $\beta$-parameter is not a fixed universal parameter, but should run with both the energy and the external magnetic field whenever we go beyond the tree-level. Moreover, in a BI ED coupled to the matter sector, as in the electroweak case with a non-linear realization of the hypercharge U(1)-factor, non-perturbative effects must be included such that the choice $\sqrt{\beta} \gtrsim 100$ GeV may be justified. We would to finally point out that in the work of ref. \cite{Tseytlin}, the reader may find the microscopic origin of the Born-Infeld action as described in the framework of string theory. The microscopic origin of the CFJ term, whose effects we are inspecting in presence of the Born-Infeld Lagrangian, can also backed in a string scenario \cite{LSVstring theory}.
The coefficients of the expansion around the magnetic background are
\begin{eqnarray}\label{cBI}
d^{BI}=\frac{1}{\beta^2+B^2}
\hspace{0.4cm} \mbox{and} \hspace{0.4cm}
f^{BI}= \frac{1}{\beta^{2}} \; ,
\end{eqnarray}
in which both the coefficients go to zero when $\beta \rightarrow \infty$. Substituting these results in (\ref{Deltani}), the solutions for the birefringence in the BI theory (when $g^2 \, B \ll 1$) are given by
\begin{subequations}
\begin{eqnarray}
\Delta n_{11}^{(BI)} \!&=&\! 1-\sqrt{1+\frac{B^2}{\beta^2}}  \; ,
\label{Deltan1BI}
\\
\Delta n_{22}^{(BI)}(k) \!&\simeq&\! \sqrt{\frac{B^2+\beta^2}{B^2+\beta^2+v^2\beta^2/k^2}}
\nonumber \\
&&
\hspace{-0.5cm}
-\frac{1}{\sqrt{1+v^2/k^2}}
\left[\sqrt{1+\frac{B^2}{\beta^2}}+\frac{ g^2\,B^2 (B^2+\beta^2)/2}{ (m^2-v^2)\beta^2+\left(k^2+m^2\right)B^2} \right]  \; ,
\label{Deltan2BI}
\\
\Delta n_{33}^{(BI)}(k) \!&\simeq&\!  \frac{g^2\,B^2}{2(k^2+m^2)^{3/2}} \frac{1}{\sqrt{1+B^2/\beta^2}}
%\nonumber \\
%&&
%\hspace{-0.5cm}
\times
%\frac{k^2}{(1+B^2/\beta^2)\,m^2-v^2}
\nonumber \\
&&
\hspace{-0.5cm}
\times
\, \frac{m^2B^2(k^2+m^2)(B^2+2\beta^2)+(m^2-v^2)(k^2+m^2+v^2)\beta^4}{[\,m^2B^2+(m^2-v^2)\beta^2\,][\,(k^2+m^2)B^2+(m^2-v^2)\beta^2\,]} \; ,
\label{Deltan3BI}
\\
\Delta n_{12}^{(BI)}(k) \!&=&\! 1-\frac{1}{\sqrt{1+v^2/k^2}}\left[ \sqrt{1+\frac{B^2}{\beta^2}}+\frac{ g^2\,B^2 (B^2+\beta^2)/2}{ (m^2-v^2)\beta^2+\left(k^2+m^2\right)B^2} \right]  ,
\label{Deltan12BI}
\\
\Delta n_{21}^{(BI)}(k) \!&\simeq&\! \sqrt{\frac{B^2+\beta^2}{B^2+\beta^2+v^2\beta^2/k^2}}-\sqrt{1+\frac{B^2}{\beta^2}} \; ,
\label{Deltan21BI}
\\
\Delta n_{13}^{(BI)} (k) \!&\simeq&\! 1-\frac{k}{\sqrt{k^2+m^2}}\left[ 1- \frac{ g^2\,B^2\beta^2/2}{ (m^2-v^2)\beta^2+\left(k^2+m^2\right)B^2}\,\sqrt{1+\frac{B^2}{\beta^2}}\, \right] ,
\label{Deltan13BI}
\\
\Delta n_{31}^{(BI)} (k) \!&\simeq&\! \frac{1}{\sqrt{1+m^2/k^2}}-\sqrt{1+\frac{B^2}{\beta^2}}
\nonumber \\
&&
\hspace{-0.5cm}
-\frac{ g^2\,B^2\beta^2/2}{ (m^2-v^2)\beta^2+m^2\,B^2}\,\sqrt{1+\frac{B^2}{\beta^2}}\frac{m^2\,k}{(k^2+m^2)^{3/2}} \; ,
\label{Deltan31BI}
\\
\Delta n_{23}^{(BI)} (k) \!&\simeq&\! \sqrt{\frac{B^2+\beta^2}{B^2+\beta^2+v^2\beta^2/k^2}}
-\frac{1}{\sqrt{1+m^2/k^2}} \left[ 1-\frac{g^2\,B^2}{2\,(m^2-v^2)}\sqrt{1+\frac{B^2}{\beta^2}} \right]
\nonumber \\
&&
\hspace{-0.5cm}
-\frac{g^2\,B^2}{2\,(m^2-v^2)}\frac{B^2\,k\,\sqrt{k^2+m^2}}{B^2(k^2+m^2)+(m^2-v^2)\beta^2} \, \sqrt{1+\frac{B^2}{\beta^2}}
\; ,
\label{Deltan23BI}
\\
\Delta n_{32}^{(BI)}(k) \!&\simeq&\! \frac{1}{\sqrt{1+m^2/k^2}}-\frac{1}{\sqrt{1+v^2/k^2}}\,\sqrt{1+\frac{B^2}{\beta^2}}
\left[1+\frac{ g^2\,B^2\,\beta^2/2}{ (m^2-v^2)\beta^2+\left(k^2+m^2\right)B^2} \right]
\nonumber \\
&&
\hspace{-0.5cm}
-\frac{g^2\,B^2\,\beta^2/2}{B^2(k^2+m^2)+(m^2-v^2)\beta^2} \, \sqrt{1+\frac{B^2}{\beta^2}} \frac{m^2\,k}{(k^2+m^2)^{3/2}}
\; .
\label{Deltan32BI}
\end{eqnarray}
\end{subequations}
The limit $\beta \rightarrow \infty$ recovers the results (\ref{Deltan1noED})-(\ref{Deltan32noED}).
For a weak magnetic background, {\it i. e.}, $\beta \gg B$, the birefringence effect is residual in $B^2/\beta^2$
and also depends on the $g^2$ coupling constant, for $\Delta n_{ii}^{(BI)}$ :
\begin{subequations}
\begin{eqnarray}
\Delta n_{11}^{(BI)} \!&\simeq&\! -\frac{B^2}{2\beta^2}  \; ,
\label{Deltan1BIBweak}
\\
\Delta n_{22}^{(BI)}(k) \!&\simeq&\! -\frac{1}{\sqrt{1+v^2/k^2}}\left[ \frac{B^2}{2\beta^2}+\frac{g^2\,B^2}{2(m^2-v^2)} \right] \; ,
\label{Deltan2BIBweak}
\\
\Delta n_{33}^{(BI)}(k) \!&\simeq&\!  \frac{g^2\,B^2\,k}{2(k^2+m^2)^{3/2}} \frac{k^2+m^2+v^2}{m^2-v^2} \; .
\label{Deltan3BIBweak}
\end{eqnarray}
\end{subequations}
In the limit $B \rightarrow 0$, $\Delta n_{ii}^{(BI)}=0$ in the results (\ref{Deltan1BI})-(\ref{Deltan3BI}).
For the propagation effects in a low energy scale, we use $\sqrt{\beta}=16$ MeV
associated with the electron's self-energy in BI ED. In this case, the solution (\ref{Deltan1BIBweak}) has the numeric value
\begin{eqnarray}
\frac{|\Delta n_{11}^{(BI)}|}{B^2}\simeq 3.2 \times 10^{-24} \, \mbox{T}^{-2} \; ,
\end{eqnarray}
that is the same order of the PVLAS-FE experiment. Notice also that, in the limit $v \rightarrow 0$ and considering $B^2/\beta^2 \approx 0$,
the expressions $\Delta n_{22}^{(BI)}$ and $\Delta n_{33}^{(BI)}$ in (\ref{Deltan2BIBweak})-(\ref{Deltan3BIBweak}) are reduced to the result obtained in the ref. \cite{nossoJHEP}, when $k \gg m \gg v$ in the PVLAS-FE experiment :
\begin{eqnarray}
|\Delta n_{22}^{(BI)}| \simeq \Delta n_{33}^{(BI)} \simeq \frac{g^2B^2}{2m^2} \; .
\end{eqnarray}
When $B \rightarrow 0$, the variations of $\Delta n_{ij}^{(BI)} (i\neq j)$ are reduced to (\ref{Deltan12EHB0})-(\ref{Deltan23EHB0}),
that confirms the birefringence depending on the CFJ $v$-parameter.

%depend only on the axion mass and $v$-parameter :
%
%\begin{equation}
%\Delta n_{23}^{(BI)} \stackrel{B\rightarrow 0}{\simeq}
%\frac{ \sqrt{2} \, k}{ \sqrt{2k^2+m^2+v^2-|m^2-v^2|} }
%\nonumber \\
%-\frac{ \sqrt{2} \, k}{ \sqrt{2k^2+m^2+v^2+|m^2-v^2|} } \simeq \frac{|m^2-v^2|}{2k^2} \; ,
%\end{equation}
%
%if $k^2 \gg (m^2,v^2)$.
%

%
\item The modified Maxwell (ModMax) ED is set by the Lagrangian
\begin{eqnarray}\label{ModMaxL}
{\cal L}_{MM}({\cal F},{\cal G})=\cosh\gamma \, {\cal F} + \sinh\gamma \sqrt{{\cal F}^2+{\cal G}^2} \; ,
\end{eqnarray}
where $\gamma$ is a real and positive parameter of this theory. In the limit $\gamma \rightarrow 0$,
the ModMax Lagrangian reduces to the Maxwell ED. This non-linear ED has been well motivated in the
literature due to the conformal invariance. Thus, it is the only non-linear ED that preserve the
duality and the conformal symmetries in the same Lagrangian \cite{Bandos}.
The coefficients of the expansion in the magnetic background, in this case, are
\begin{eqnarray}\label{coefficientsMM}
d^{MM}=0
\hspace{0.4cm} \mbox{and} \hspace{0.4cm}
f^{MM}= 2 \, e^{\gamma} \, \frac{\sinh\gamma}{B^2} \; .
\end{eqnarray}
Thus, the variation of the refractive index for a weak axion-coupling constant are read below :
\begin{subequations}
\begin{eqnarray}
\Delta n_{11}^{(MM)} \!&=&\! 0  \; ,
\label{Deltan1MM}
\\
\Delta n_{22}^{(MM)}(k) \!&\simeq&\!
\frac{ e^{\gamma }}{\sqrt{e^{2 \gamma }+v^2/k^2}}
-\frac{e^{\gamma }}{\sqrt{1+v^2/k^2}}\left[1+\frac{e^{\gamma} \, g^2\,B^2/2}{e^{2 \gamma } \left(k^2+m^2\right)-k^2-v^2} \right]
\, , \;\;
\label{Deltan2MM}
\\
\Delta n_{33}^{(MM)}(k) \!&\simeq&\!  \frac{1}{(1+m^2/k^2)^{3/2}} \frac{m^2-v^2}{e^{2\gamma}\,m^2-v^2}
%\nonumber \\
%&&
%\hspace{-1.0cm}
%\times \;
\frac{e^{\gamma} \, g^2\,B^2/2}{e^{2\gamma}\left(k^2+m^2\right)-k^2-v^2 } \; ,
\label{Deltan3MM}
\\
\Delta n_{12}^{(MM)}(k) \!&\simeq&\! 1-\frac{e^{\gamma}}{\sqrt{1+v^2/k^2}}\left[1+\frac{e^{\gamma} \, g^2\,B^2/2}{e^{2 \gamma } \left(k^2+m^2\right)-k^2-v^2} \right] \; ,
\label{Deltan12MM}
\\
\Delta n_{21}^{(MM)}(k) \!&\simeq&\! \frac{e^{\gamma}}{\sqrt{e^{2\gamma}+v^2/k^2}}-1  \; ,
\label{Deltan21MM}
\\
\Delta n_{13}^{(MM)}(k) \!&\simeq&\! 1-\frac{1}{\sqrt{1+m^2/k^2}}\left[1-\frac{e^{\gamma} \, g^2\,B^2/2}{m^2-v^2+2e^{\gamma}\sinh(\gamma)(k^2+m^2)}\right] \; ,
\label{Deltan13MM}
\\
\Delta n_{31}^{(MM)}(k) \!&\simeq&\! \frac{1}{\sqrt{1+m^2/k^2}}-1-\frac{e^{\gamma} \, g^2 \, B^2/2}{e^{2\gamma}\,m^2-v^2}\frac{m^2\,k}{(k^2+m^2)^{3/2}} \; ,
\label{Deltan31MM}
\\
\Delta n_{23}^{(MM)}(k) \!&\simeq&\! \frac{e^{\gamma}}{\sqrt{e^{2\gamma}+\,v^2/k^2}}
\nonumber \\
&&
\hspace{-0.5cm}
-\frac{1}{\sqrt{1+m^2/k^2}}\left[1-\frac{e^{\gamma} \, g^2\,B^2/2}{m^2-v^2+2e^{\gamma}\sinh(\gamma)(k^2+m^2)} \right] \; ,
\label{Deltan23MM}
\\
\Delta n_{32}^{(MM)}(k) \!&\simeq&\!  \frac{1}{\sqrt{1+m^2/k^2}} - \frac{e^{\gamma}}{\sqrt{1+v^2/k^2}}\left[ 1+\frac{e^{\gamma} \, g^2\,B^2/2}{e^{2 \gamma } \left(k^2+m^2\right)-k^2-v^2} \right]
\nonumber \\
&&
\hspace{-0.5cm}
-\frac{e^{\gamma} \, g^2\,B^2/2}{e^{2\gamma}\,m^2-v^2} \frac{m^2\,k}{(k^2+m^2)^{3/2}}
\; .
\label{Deltan32MM}
\end{eqnarray}
\end{subequations}
The results (\ref{Deltan1noED})-(\ref{Deltan32noED}) also are recovered in the limit $\gamma \rightarrow 0$.
Notice that, with $\gamma \neq 0$, the birefringence remains in the second solution (\ref{Deltan2MM}) when
$B \rightarrow 0$ :
\begin{eqnarray}\label{Deltan22MMB0}
\Delta n_{22}^{(MM)}(k) \simeq
\frac{e^{\gamma }}{\sqrt{e^{2 \gamma }+v^2/k^2}}-\frac{e^{\gamma }}{\sqrt{1+v^2/k^2}} \; ,
\end{eqnarray}
whereas $\Delta n_{33}^{(MM)}=0$ in this limit. Again, $\Delta n_{33}^{(MM)}$ is proportional to $g^2\,B^2$,
but in the case of ModMax ED, the result constraints the axion mass and the $v$-CFJ parameter with the $\gamma$-ModMax.
The particular result (\ref{Deltan22MMB0}) is the case in which the ModMax ED is added to the CFJ term
without the presence of the axion. The result (\ref{Deltan22MMB0}) shows the birefringence solution that depends directly on the
$v$-CFJ, and on the $\gamma$-ModMax parameters. When $v\rightarrow 0$, the birefringence of (\ref{Deltan22MMB0}) is equal to (\ref{Deltan12MM}).
This is the simplest case of the pure ModMax ED.
Turning off the magnetic background, the results for $\Delta n_{ij}^{(MM)}(i \neq j)$ are reduced to
\begin{subequations}
\begin{eqnarray}
\lim_{B \rightarrow 0}
\Delta n_{12}^{(MM)}(k) \!&=&\! 1-\frac{e^{\gamma}}{\sqrt{1+v^2/k^2}} \; ,
\label{Deltan12MMB0}
\\
\lim_{B \rightarrow 0}\Delta n_{21}^{(MM)}(k) \!&=&\! \frac{e^{\gamma}}{\sqrt{e^{2\gamma}+\,v^2/k^2}}-1  \; ,
\label{Deltan21MMB0}
\\
\lim_{B \rightarrow 0}\Delta n_{13}^{(MM)}(k) \!&=&\! - \Delta n_{31}^{(MM)}(k) = 1-\frac{1}{\sqrt{1+m^2/k^2}} \; ,
\label{Deltan13MMB0}
\\
%\Delta n_{31}^{(MM)}(k) \!&\simeq&\! \frac{1}{\sqrt{1+m^2/k^2}}-1 \; ,
%\label{Deltan13MM}
%\\
\lim_{B \rightarrow 0}\Delta n_{23}^{(MM)}(k) \!&=&\! \frac{e^{\gamma}}{\sqrt{e^{2\gamma}+\,v^2/k^2}}-\frac{1}{\sqrt{1+m^2/k^2}} \; ,
\label{Deltan23MMB0}
\\
\lim_{B \rightarrow 0}\Delta n_{32}^{(MM)}(k) \!&=&\!  \frac{1}{\sqrt{1+m^2/k^2}} - \frac{e^{\gamma}}{\sqrt{1+v^2/k^2}} \; ,
\label{Deltan32MMB0}
\end{eqnarray}
\end{subequations}
where the $\gamma$-ModMax parameter plays a fundamental rule for the birefringence in the results of $\Delta n_{12}^{(MM)}$,
$\Delta n_{21}^{(MM)}$, $\Delta n_{23}^{(MM)}$ and $\Delta n_{32}^{(MM)}$. If the CFJ $v$-parameter predominates respect to the axion mass, both the $\gamma$- and $v$-parameters corroborate to the birefringence. In the case of $v \rightarrow 0$, just the $\Delta n_{12}^{(MM)}$ and $\Delta n_{32}^{(MM)}$ have the same birefringence of
$\Delta n_{22}^{(MM)}$, that is, $\Delta n_{12}^{(MM)}=\Delta n_{32}^{(MM)}=\Delta n_{22}^{(MM)}=1-e^{\gamma}$
for the pure ModMax ED, with $m\simeq0$.

%
%In the case of $\Delta n_{23}^{(MM)}$, the birefringence is null in an intense magnetic field. The variation $\Delta n_{31}^{(MM)}$
%is finite on both the limits $B \rightarrow 0$, and $B\rightarrow \infty$. When the magnetic background is intense, $\Delta n_{31}^{(MM)}=-1$,
%that does not depend on the any parameter of the theory.
%

In the PVLAS-FE experiment, the wavelength corresponds to $k=0.25$ eV in an external magnetic field of $B_{ext}=2.5$ T. The axion parameter space $(m,g)$ is quite involved with many restrictions. For instance, let us consider the upper bound $g < 6.4 \times 10^{-8} \, \mbox{GeV}^{-1}$ at $95 \, \% \, \mbox{C. L.}$, which is consistent with small axion mass $m < 10^{-3}$ eV \cite{25years}. Moreover, for the Lorentz violating background, we shall also assume the upper bound $v < 10^{-23}$ GeV \cite{Kostelecky_Russell}.
%Keeping this in mind, it is possible to consider the approximation $k \gg (m,v)$ in the results (\ref{Deltan2MM}) and (\ref{Deltan3MM}).
Using that $\Delta n_{PVLAS-FE}/B^2=(19\pm 27) \times 10^{-24}\,\mbox{T}^{-2}$, the term that depends on the magnetic background
in (\ref{Deltan2MM}) can be neglected, in which the $\gamma$-parameter is constrained by the relation :
\begin{subequations}
\begin{eqnarray}
|1-e^{\gamma}|
%-\frac{g^2\,B^2}{4k^2} \, \frac{e^{\gamma}}{\sinh(\gamma)}
& < & 1.18 \times 10^{-22} \; ,
\label{Deltan2MMapprox}
%\\
%\frac{g^2 \, B^2}{4k^2\sinh(\gamma)} \frac{m^2-v^2}{e^{2\gamma}\,m^2-v^2} & < & 1.18 \times 10^{-22} \; ,
%\label{Deltan3MMapprox}
\end{eqnarray}
\end{subequations}
%
%for $\Delta n_{22}^{(MM)}$ and $\Delta n_{33}^{(MM)}$, respectively. From the eq. (\ref{Deltan2MMapprox}) with $g < 6.4 \times 10^{-8} \, \mbox{GeV}^{-1}$,
that for $\gamma \ll 1$, we arrive at the following constraint for the $\gamma$-ModMax parameter
\begin{eqnarray}
|\gamma| < 1.18 \times 10^{-22} \; ,
\end{eqnarray}
which is compatible with the $\gamma$-result obtained in ref. \cite{Sorokin}.

\end{enumerate}
\section{Conclusions and perspectives}
\label{sec5}
We propose a general non-linear electrodynamics coupled to a scalar axion to which we adjoin the Carrol-Field-Jackiw (CFJ) term. We expand the Lagrangian of the model around a uniform electromagnetic background field up to second order in the photon propagation field. The CFJ term introduces a background $4$-vector $v^{\mu}=(v^{0},{\bf v})$, that consequently, breaks the Lorentz symmetry in the theory. The case with only a uniform magnetic background field $({\bf B})$ is analyzed, where the properties of the wave propagation are discussed. Thereby, we calculate the dispersion relations of the model for a space-like $(v^{0}=0)$ CFJ term. The wave propagation is affected by three vectors ${\bf B}$, ${\bf k}$ (wave vector) and ${\bf v}$. The dispersion relations are obtained for two cases : (a) when ${\bf B}$, ${\bf k}$ and ${\bf v}$ are perpendicular among themselves, (b) when ${\bf v}$ is perpendicular to ${\bf B}$ and ${\bf k}$, but ${\bf B}$ and ${\bf k}$ are parallel vectors. These results allow us to defined the refractive index of this medium, and posteriorly, we discuss the birefringence phenomenon under these conditions. Since there are three different solutions for the dispersion relations, we discuss the possible cases of birefringence, where the variation of the refractive index in the medium is $\Delta n_{ij}=n_{i\parallel}-n_{i\perp}$, with $i,j=1,2,3$.  We apply the birefringence results for three cases of well-known non-linear ED in the literature : Euler-Heisenberg, Born-Infeld, and the ModMax ED. When the non-linearity is null, the birefringence effect emerges due to the axion coupling constant and the presence of the magnetic background. In some situations, when the magnetic field is turned off, the birefringence is due to the CFJ parameter, the axion mass and the parameter of the non-linear ED.

One of the solutions of Euler-Heisenberg ED exhibits the birefringence result  $\Delta n_{11}^{(EH)}/B^2 \simeq  5.3 \times 10^{-24} \, \mbox{T}^{-2}$, that is compatible with the PVLAS-FE experiment for vacuum magnetic birefringence, {\it i. e.}, $\Delta n_{PVLAS-FE}/B^2=(19\pm 27) \times 10^{-24}\,\mbox{T}^{-2}$. The third solution (\ref{Deltan3EH}) shows positive birefringence as function of the background magnetic field. In the case of the Born-Infeld ED, one of solutions for the birefringence yields $|\Delta n_{11}^{(BI)}|/B^2\simeq3.2 \times 10^{-24}\,\mbox{T}^{-2}$, when the Born-Infeld parameter is bounded by the finite electron's self-energy. This numeric value is of the same order of the value found in the PVLAS-FE experiment. The result of $\Delta n_{33}$ is proportional to $g^2\,B^2$ (axion coupling squared times the magnetic background field) in all the non-linear EDs analyzed in this paper. In the case of the ModMax ED, the birefringence described by $\Delta n_{22}^{(MM)}$ constraints the ModMax parameter $(\gamma)$. When the solutions of $\Delta n_{ij}^{(MM)}$, for $i \neq j$, are analyzed, the $v$- and $\gamma$-parameters play a fundamental rule in the case of $\Delta n_{12}^{(MM)}$, $\Delta n_{21}^{(MM)}$, $\Delta n_{23}^{(MM)}$ and $\Delta n_{32}^{(MM)}$, in the ModMax ED. When the magnetic background is turned off (that is equivalent to consider $g\rightarrow0$), and for a massless axion, the birefringence emerges thanks to the $v$- and $\gamma$-parameters. In the simplest case in which $v\simeq0$ and $\gamma \neq 0$, the birefringence in pure ModMax is recovered in the results $\Delta n_{12}^{(MM)}=\Delta n_{32}^{(MM)}=\Delta n_{22}^{(MM)}=1-e^{\gamma}$, whereas the others contributions are nulls. Using the PVLAS-FE experiment result in $\Delta n_{22}^{(MM)}$, and the axion-coupling $g < 6.4 \times 10^{-8} \, \mbox{GeV}^{-1}$ (with $95 \, \% \, \mbox{C. L.})$, we obtain the upper bound $\gamma < 1.18 \times 10^{-22}$.

In addition, from the dispersion relations, we also show that the parameters of non-linear ED, axion sector and LSV can combine together to generate effective masses for the photon and axion fields. In connection with this special issue, let us recall that the stars - and the Sun, in particular - with their large-scale magnetic fields are important astrophysical sources of axions and low-mass ALPs, mainly through the Primakoff Effect. The photon-axion conversion probability after traveling a distance in presence of a uniform magnetic field is sensible to the axion mass. Now, due to the combination of effects as we are inspecting in this paper, we expect that our expressions for the photon and axion effective masses may be used to set bounds on the various parameters in future Helioscope and dark matter experiments.

In our purpose of investigating how different new physics interfere with one another through the photon sector, we point out that, in an interesting recent article, Li and Ma \cite{Hao LI} pursue an inspection on the effects stemming from Loop Quantum Gravity (LQG) corrections to both the photon and fermionic matter sectors of Electrodynamics. Among these corrections, there appears a non-linear (actually, cubic) term in the extended Ampère-Maxwell equation. Though modulated by LQG parameters, very strong external magnetic fields at the astrophysical or those generated in relativistic heavy ion colliders may be sufficient to enhance the associated LQG effects and, therefore, one can compute how these latter effects contribute to the axion physics through the photon-axion coupling, as we have considered here.
Finally, considering still our motivation to relate non-linear photon effects with axion physics, we recall that we have here considered as electromagnetic backgrounds only constant and uniform fields. It remains to be contemplated, for example, situations with non-uniform external electric/magnetic fields that will be exchanging energy and momentum with the photon-axion system, and to compute the modified dispersion relations, the corresponding group velocities, refractive indices and birefringence which will become space-dependent as a consequence of the non-uniformity of the background.

%%%%%%%%%%%%

{\bf Acknowledgments}:
The authors are grateful to Gustavo P. de Brito for his valuable suggestions on this paper.
L.P.R. Ospedal expresses his gratitude to FAPERJ (grant number E-26/203.997/2022) for his postdoctoral fellowship. J.M.A Paixão is grateful to the National Conselho Nacional de Desenvolvimento Científico e Tecnológico (CNPq) for supporting his work.

\end{document}